\documentclass[a4paper,12pt]{article}
\usepackage[warn]{mathtext}
\usepackage[T2A]{fontenc}
\usepackage[cp1251]{inputenc}
\usepackage[english]{babel}
\usepackage{amssymb,amsfonts,amsmath,mathtext,cite,enumerate,float}
\usepackage{graphicx}
\usepackage{indentfirst}

\makeatletter
\renewcommand{\@biblabel}[1]{#1.}
\makeatother

\usepackage{geometry}
\begin{document}
\begin{center}
\textbf{Stability of a Nonwetting Liquid in a Nanoporous Medium}
\end{center}

\begin{center}
\textbf{V.D. Borman, A.A. Belogorlov, V.A. Byrkin, V.N. Tronin, V.I. Troyan}
\end{center}

Department of Molecular Physics, National Research Nuclear University MEPhI, Kashirskoe sh. 31, Moscow 115409, Russia

\begin{abstract}
A physical mechanism has been proposed to describe the formation of a stable state of a nonwetting liquid after filling of a porous medium at an increased pressure with the subsequent reduction of the overpressure to zero. It has been shown that the observed transition of the system of nonwetting-liquid nanoclusters to the stable state in a narrow range of filling factors and a narrow temperature interval is due to a decrease in the surface free energy without change in the chemical potential owing to the collective interactions between liquid nanoclusters in the neighboring pores. This effectively corresponds to their mutual attraction. The model makes it possible to describe the observed dependences of the volume of the liquid in pores in the stable state on the filling factor and temperature.
\end{abstract}

To interpret the data on the characterization of the porous medium by means of mercury porosimetry \cite{1,2,3,4}, it is necessary to understand the mechanisms of phenomena and processes occurring when a porous medium is filled with a nonwetting liquid. One of these phenomena is entrapment, i.e., the formation of a stable state of the nonwetting liquid filling the medium at the increased pressure with a subsequent reduction of the overpressure to zero. Information on the distribution of the confined dispersed liquid over pores of various sizes is also necessary for, e.g., the interpretation of investigations of the dependence of the melting temperature on the size of clusters in the case of entrapment \cite{5,22}. The stability of the dispersed nonwetting liquid in the porous medium was observed in a number of works studying the intrusion of mercury into porous glasses \cite{4,5}, as well as the intrusion of water and aqueous solutions of salts \cite{6,7,8,9,10,11} and organic materials \cite{4,12,13,14,15} into various hydrophobic silica gels, structured porous media MCM-41, and zeolites. It was found that the fraction of the volume of a liquid confined in a medium can vary from 0 to 1 in media with different porosities and depends on the average radius, width of the pore size distribution, concentration of substances in water, and temperature. As far as we know, the stability of a dispersed nonwetting liquid in a porous medium has not yet been systematically studied. For example \cite{4,16,17} that this phenomenon can be due to the structural features of the medium: the pore size distribution, the barrier for the formation of the liquid--vapor interface at extrusion from a filled cylindrical pore, and hysteresis of phenomenological wetting angles at intrusion--extrusion. Analysis indicates that all mentioned data cannot be described under these assumptions. 

In \cite{18}, it was shown that the fraction of a liquid in a stable state depends on the filling factor and the transition of a part of the liquid to the stable state occurs in a narrow range of the filling factor and in a narrow temperature range.

In this work, taking into account both the percolation properties of the medium with pores of various sizes and the ``multiparticle interaction'' between liquid clusters in neighboring interconnected pores, we propose a mechanism of the appearance of a stable state of a nonwetting liquid in a porous medium. The energy of the ``interaction'' between two liquid clusters in neighboring pores connected through a mouth (throat) is defined as a change in the surface energy of the liquid in the mouth after the appearance (disappearance) of a liquid cluster in the neighboring pore. The stable state of the nonwetting liquid appears as a result of a dispersion transition when filling factors and temperatures are such that the energy of the ``multiparticle interaction'' (attraction) between neighboring liquid clusters is higher than the energy of the liquid--solid interface. The transition to the stable state is described in the framework of the analytical percolation theory.

\begin{figure}[h]
\begin{minipage}[h]{0.49\linewidth}
\center{\includegraphics[width=1\linewidth]{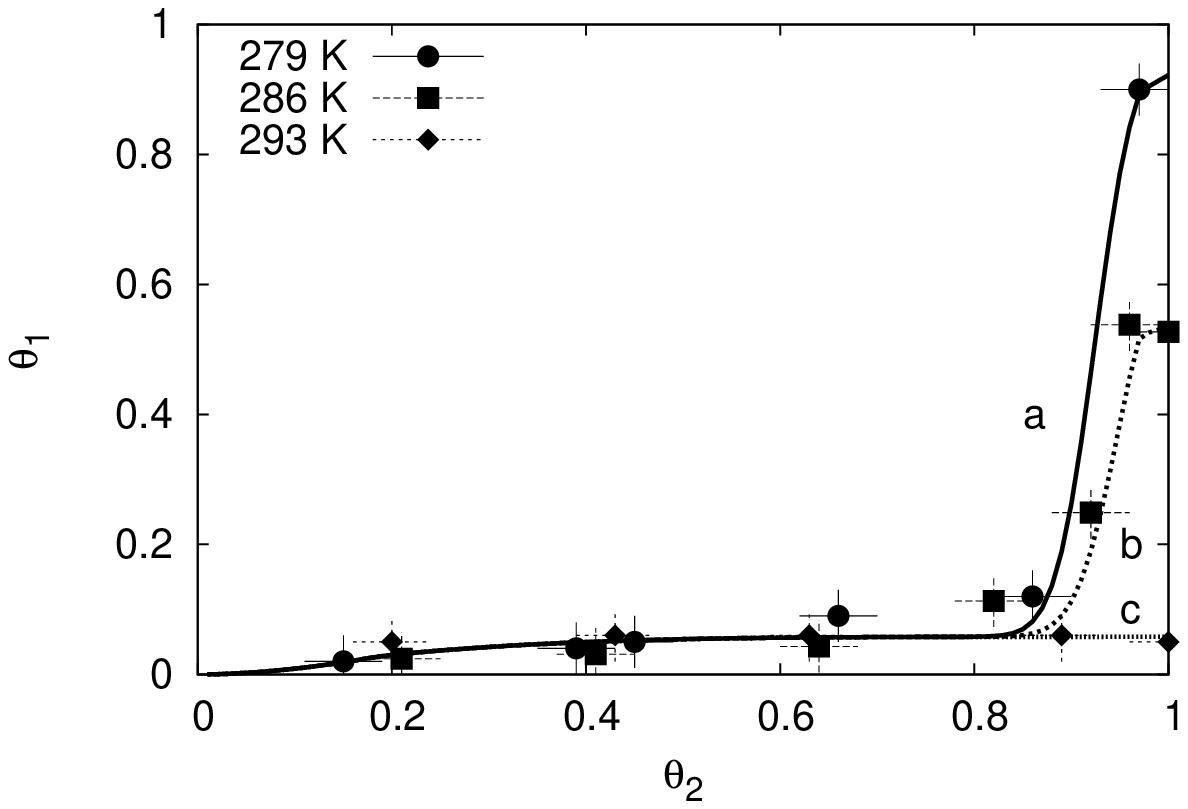} \\ (a)}
\end{minipage}
\hfill
\begin{minipage}[h]{0.49\linewidth}
\center{\includegraphics[width=1\linewidth]{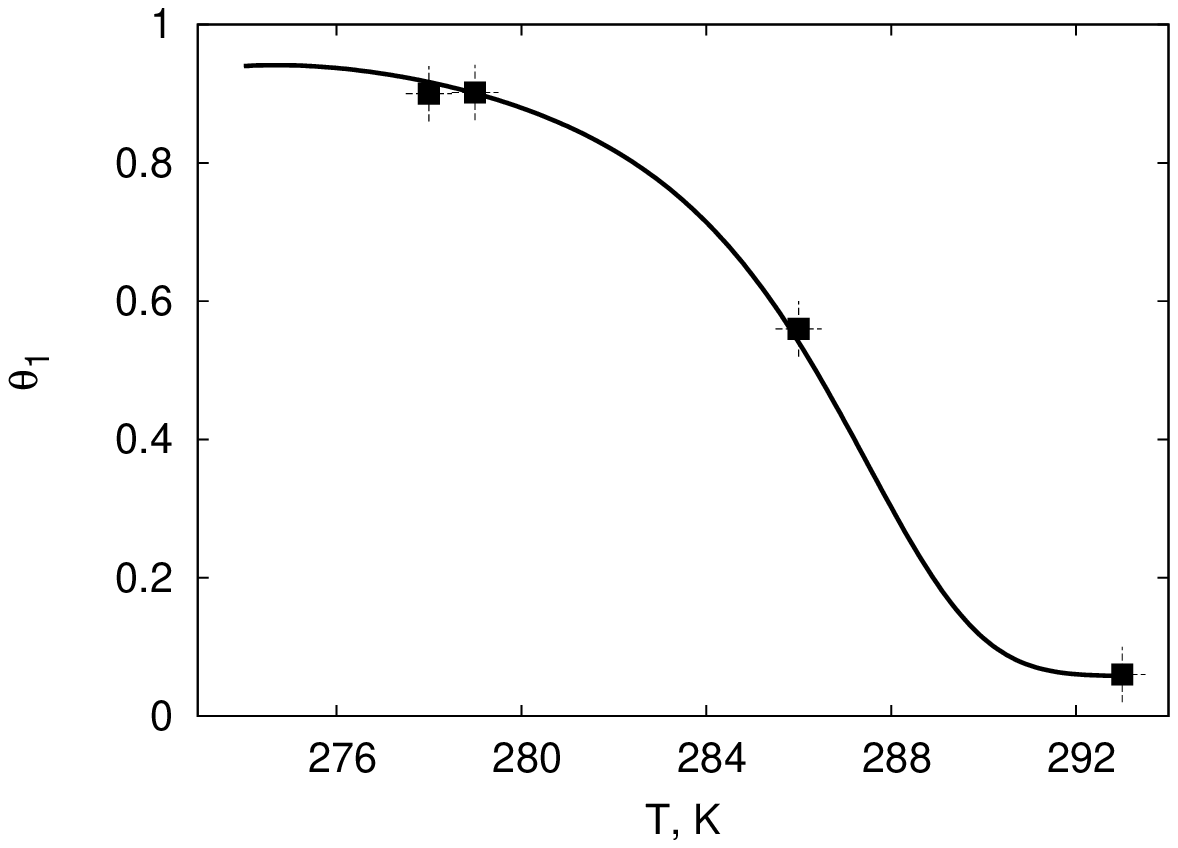} \\ (b)}
\end{minipage}
\caption{(a) Dependences $\theta _{2}$ ($\theta _{1}$) for three temperatures $T= 279$, 286, and 293 K and (b) dependence $\theta _{2}(T)$ at $\theta _{1} =1.0$ (experimental points in comparison with the calculations using the model proposed in this work shown by the lines)}
\label{ris:image1}
\end{figure}

The fraction $\theta_2$ of the volume of water remaining in the L23C8 porous medium after the first intrusion cycle as obtained in \cite{18} is shown in Fig.~\ref{ris:image1} as a function of (a) the fraction $\theta_1$ of the volume of pores filled in the first intrusion--extrusion cycle for three temperatures 279, 286 and 293 K and (b) the temperature at $\theta _{1} =1.0$.


According to Fig.~\ref{ris:image1}, the fraction of the volume of the remaining liquid at $\theta _{1} >0.16$ is $\theta _{2} \sim 0.06$ at $T=293$ K. A decrease in the temperature to 286 K does not change the remaining volume at $\theta _{1} < 0.85$, but leads to an increase in the fraction of the confined liquid at $\theta _{1} > 0.85$ ($\theta _{2} \approx 0.6$ at $\theta _{1} =1$). With a further reduction of the temperature to 279 K, the fraction of the confined volume remains unchanged at $\theta _{1} <0.85$ and a further increase in the fraction of the remaining liquid is observed at $\theta _{1} >0.85$. Thus, the dependence of the fraction of the volume of the remaining liquid on both the fraction $\theta {}_{1} $ of the volume of the pores filled by the liquid and the temperature of the system has a threshold behavior. It follows from Fig.~\ref{ris:image1} that the threshold values of the temperature and fraction of the filled volume are $T_c \approx 287$ K and $\theta _{c0} \approx 0.85$, respectively. In this work, we propose a model that takes into account both the properties of the interface and the percolation properties of the space of the pores in the medium and makes it possible to describe the observed phenomenon.

We assume that a porous body is formed by a solid framework in which pores form a spatial structure as overlapping spheres distributed in the radius. This model of the porous medium is a generalization of the model of randomly located spheres, which is widely used to describe porous media \cite{19}. In the model of randomly located spheres, an elementary pore is represented as a spherical cavity with notches (throats). In the model under consideration, a ``quantum'' of a change in the volume of the liquid in the medium at quasistatic intrusion (extrusion) is the volume of one pore. It is assumed that the volume of the throats is negligibly small compared to the volume of the pores. When the liquid is intruded into a pore and is extruded from it, liquid menisci are formed it these throats. We suppose that the width of the pore size distribution $\delta R$ satisfies the inequality $\delta R/\bar{R}<3$, which guarantees that the spread of the percolation of the threshold in interconnected pores is independent of the radius of pores \cite{20}. 

Pores can be filled only when they are connected with each other through throats and with the surface of the porous medium. This occurs if porosity is such that the system of pores in it is above the percolation threshold, i.e., $\varphi >\varphi _{c} $, where $\varphi _{c} =0.16 \div 0.3$ is the percolation threshold \cite{19, 21}, which characterizes the porous medium. Pores are connected with each other because an infinite (geometric) cluster, which consists of interconnected pores, appears at $\varphi =\varphi _{c} $. At $\varphi >\varphi _{c} $, the intrusion of the liquid into the disordered porous medium is the intrusion of the liquid into the percolation cluster consisting of interconnected pores of various sizes. Only pores with the radius larger than a certain value can be filled at a given pressure. 

The extrusion of the liquid from a pore in such a medium with a decrease in the overpressure can be considered as a result of the extrusion of the liquid from a pore belonging to the percolation cluster of interconnected pores of various sizes filled with the liquid. The liquid can be extruded from the pore under two conditions. First, its extrusion should be energetically profitable; this means that negative work should be spent on the extrusion of the liquid from the pore owing to a change in the energy of the porous medium--liquid interface and the formation of menisci in the throats of the neighboring pores. Second, since the liquid can be extruded only through the connected system of filled pores, geometrical paths should exist for its extrusion from a given pore. 

\begin{figure}[H]
\center{\includegraphics[width=0.9\linewidth]{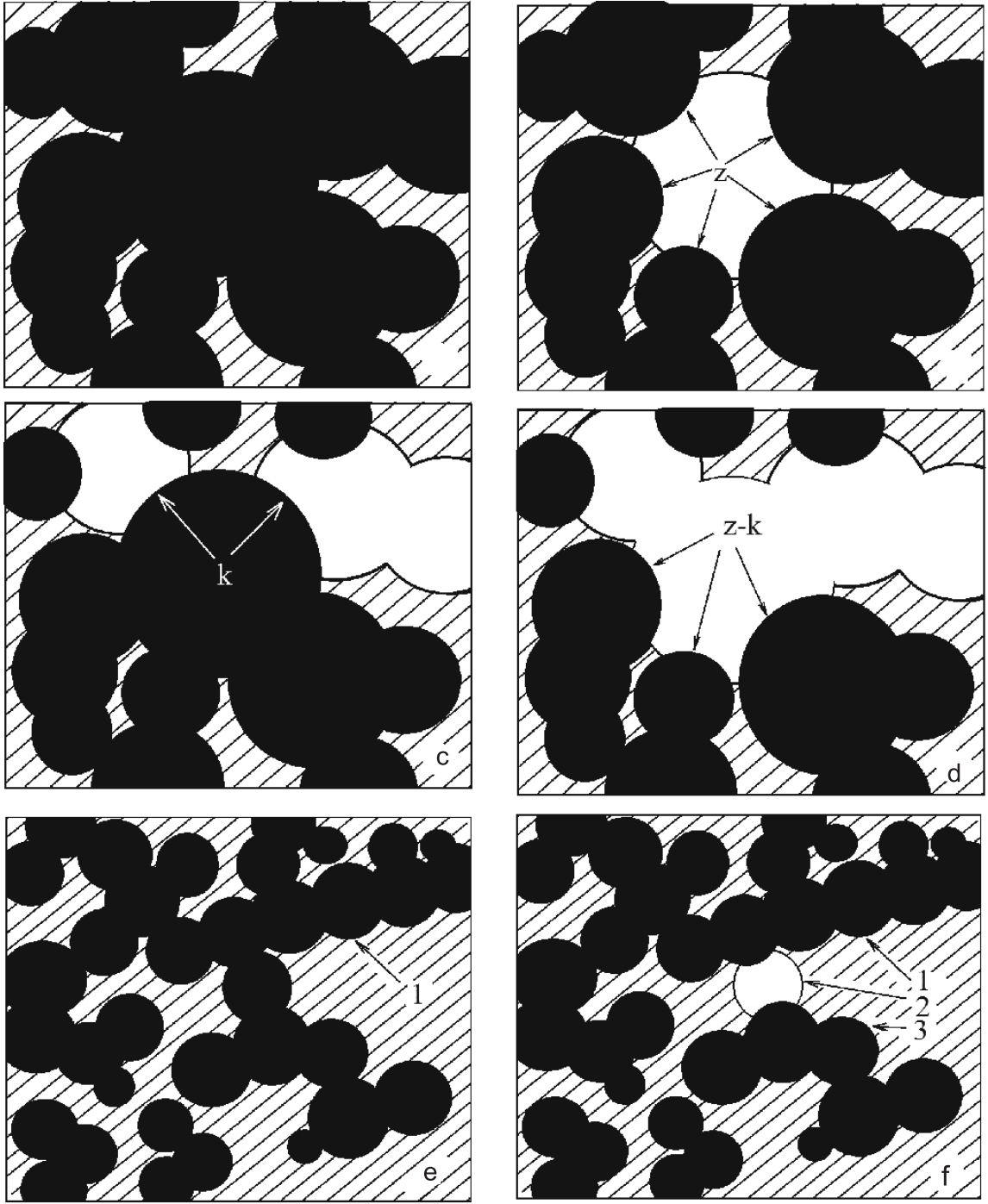}}
\caption{Schematic representation of a change in the state of the porous medium and the number of menisci in the pore before and after its extrusion of the liquid from it at the filling factors of the porous medium (a, b) $\theta _{1}=1$ and (c, d) $\theta _{1}<1$ and (d, e) schematic representation of the cluster of filled pores in the absence of paths for the extrusion of the liquid}
\label{ris:image2}
\end{figure}

Entrapment of the liquid in such a medium can be due to two reasons: (i) geometrical, when clusters containing a finite number of filled pores without paths for extrusion are formed in the porous medium and (ii) energetic, when the liquid in pores remains in the stable state with a decrease in the overpressure to zero because of the positive work necessary for the extrusion of the liquid. Figure~\ref{ris:image2} shows the schematic representation of a change in the state of the porous medium and the number of menisci in the pore before and after extrusion at various filling factors. Figures~\ref{ris:image2}a, \ref{ris:image2}c, and \ref{ris:image2}e show the states before the extrusion of the liquid from the pore and Figs.~\ref{ris:image2}b, \ref{ris:image2}d, and \ref{ris:image2}f show the states after extrusion of the liquid from the pore. Black color indicates filled pores, white color marks empty pores, and shading indicates the framework of the porous medium. Arrows in Figs.~\ref{ris:image2}a--\ref{ris:image2}d indicate menisci and their numbers $z$ and $k$. In Figs.~\ref{ris:image2}e and \ref{ris:image2}f, 1 is the ``infinite'' cluster of filled pores, 2 is the pore from which the liquid is extruded, and 3 is the formed cluster of filled pores extruded from which is impossible because of the break of connection with the ``infinite'' cluster of filled pores.

We will analyze successively all reasons for entrapment. To qualitatively analyze the energy reason for entrapment, we calculate a change in the energy of the system at the disappearance of a liquid nanocluster in one of the pores in the completely filled medium at $\theta _{1} =1$, which consists of pores of the same radius, when the overpressure vanishes (Figs.~\ref{ris:image2}a, \ref{ris:image2}b). Following \cite{22}, we accept that the chemical potential of the liquid in pores with a size of 13 nm does not change. In this case, a change in the energy at the extrusion of the liquid from a pore is determined by the difference between the surface energies after ($E_{o} $) and before ($E_{i} $) the extrusion of the liquid. Energy $E_{i} $ is the energy of the liquid--solid interface in the pore with throats the number of which is equal to the number $z$ of the nearest neighbors: $E_{i}=\sigma _{ls} (S-zS_{1})$, where $S$ is the area of the pore, $S_{1} $ is the area of the throats, and $\sigma _{ls} $ is the solid--liquid surface tension coefficient. Energy $E_{0} $ is the sum of the energy of the solid--gas interface $\sigma _{sg} (S-zS_{1} )$ ($\sigma _{sg} $ is the solid--gas surface tension coefficient) and the energy of menisci formed in the throats of the pore $\sigma _{\lg } zS_{1} $ ($\sigma _{\lg } $ is the liquid--gas surface tension coefficient, Figs.~\ref{ris:image2}a, \ref{ris:image2}b). Thus, a change in the energy of the system at the extrusion of the liquid from the pore in the completely filled porous body $\Delta E(\theta _{1} =1)$ is given by the expression

\begin{equation}
\label{eq1} 
\Delta E(\theta _{1} =1)=\sigma _{\lg } zS_{1} +(\sigma _{sg} -\sigma _{ls} )(S-zS_{1} ).    
\end{equation}

For the liquid that does not wet this porous body, we have $\sigma _{sg} <\sigma _{ls} $; for this reason, the extrusion of the liquid from the pore is energetically profitable if $\Delta E(\theta =1)=\sigma _{\lg } zS_{1} +(\sigma _{sg} -\sigma _{ls} )(S-zS_{1} )<0$. Correspondingly, if $\sigma _{\lg } zS_{1} >(\sigma _{ls} -\sigma _{sg} )(S-zS_{1} )$, the liquid is not extruded from the pore in the completely filled porous body. Estimates of $\Delta E$ show that entrapment is possible because the energy spent on the creation of menisci in the throats of the neighboring pores is higher than the energy released at the disappearance of the porous body--nonwetting liquid interface.

In a partially filled porous medium, $\theta _{1} <1$ and a filled pore in the initial state is surrounded by both the filled and empty pores (Fig.~\ref{ris:image2}c) with menisci in their throats. When the liquid is extruded from such a pore, these menisci disappear, but new menisci are formed (Fig.~\ref{ris:image2}d). If the pore in the initial state has $k$ menisci, the number of menisci in the final state is $z-k$. As a result, a change in the energy of the system at the extrusion of the liquid from the pore in the partially filled porous medium is

\begin{equation} 
\label{eq2} 
\Delta E(\theta _{1} )=\sigma _{\lg } (z-2k)S_{1} +(\sigma _{sg} -\sigma _{ls} )(S-zS_{1} ).    
\end{equation}

Therefore, if $\sigma _{\lg } (z-2k)S_{1} >(\sigma _{ls} -\sigma _{sg} )(S-zS_{1} )$, the liquid nanocluster is stable and the liquid is not extruded from the pore in the partially filled porous body.

According to Eq.~(ref{eq2}), the condition of entrapment depends on the number of empty pores $k$ around the filled pore. A randomly chosen pore in the partially filled porous medium is in contact with $k$ empty pores and, therefore, is surrounded by $z-k$ filled pores. We assume that $\frac{z-k}{z} \approx \theta _{1} $ on average; in this case, $k\approx (1-\theta _{1} )z$. The condition of entrapment of the liquid in the pore becomes

\begin{equation}
\label{eq3} 
\sigma _{\lg } (2\theta _{1} -1)\eta >(\sigma _{ls} -\sigma _{sg} )(1-\eta ), \eta =\frac{S_{m} }{S} ,   
\end{equation} 

where $\eta $ is the connectivity factor, which is the ratio of the area of menisci average in the environment to the pore surface area $S$, and $S_{m} $ is the area of all menisci of the pore. In the general case, $S_{m} $ is the area of the menisci per pore in all throats of surrounding pores.

Under the accepted assumptions, inequality (\ref{eq3}) is not satisfied at $\theta _{1} <1/2$. Consequently, entrapment is possible only when the filling factor of the porous medium is $\theta _{1} >\theta _{c0} \approx 1/2$. At $\theta _{1} <\theta _{c0} $, the number of menisci in the initial state is larger than that in the final state, and the process in which these menisci disappear and the number of menisci newly appearing in the neighboring pores at the extrusion of the liquid is smaller than the number of disappearing menisci is energetically profitable. Thus, the state of the (nonwetting liquid--nanoporous body) system is metastable at $\theta _{1} <\theta _{c0} $; i.e., after removal of the overpressure, the extrusion of the liquid from the porous medium is energetically profitable. At $\theta _{1} >\theta _{c0} $, the stable state of the liquid is a state in which stable liquid nanoclusters are present in pores. In the general case, $\theta _{c0} $ is determined by the structure of the porous medium, in particular, depends on the pore size distribution. 

A change in the energy of the system at the extrusion of the liquid from a randomly chosen filled pore (the disappearance of the liquid nanocluster in the porous medium) after removal of the overpressure can be calculated using analytical percolation theory methods \cite{23}. To this end, it is necessary to calculate a change in the energy of menisci average over all possible configurations (mutual arrangements) of pores of various sizes. We perform this calculation in the mean field approximation for the geometrical state of the medium with the infinite cluster of filled pores. A filled pore with a liquid nanocluster has neighbors, and contacts (interactions) between nanoclusters occur in the throats of pores, where a meniscus appears. Energy $\Delta E$ that should be spent on the extrusion of the liquid from the pore with radius $R$ in the medium with filled pores of radii $\{ R_{i} \} $ at pressure $p$ depends on the filling factor of the porous medium and radii $R$ and $\{ R_{i} \} $. This energy is the sum of the energy of the porous medium--liquid interface $\delta \varepsilon _{1} (R,\{ R_{i} \} )$ and the energy necessary for the formation of menisci in the throats of the neighboring pores $\delta \varepsilon _{int} (R,\{ R_{i} \} ,\theta _{1} )$, which is due to the interaction of the liquid nanocluster in the pore with liquid clusters in surrounding pores:

\begin{equation}
\label{eq4}
\begin{gathered}
\Delta E(R,\{ R_{i} \} ,\theta _{1} )=\delta \varepsilon _{1} (R,\{ R_{i} \} )+\delta \varepsilon _{int} (R,\{ R_{i} \} ,\theta _{1} ) \\[0.3cm]
{\delta \varepsilon _{1} (R,\{ R_{i} \} )=pV-\delta \sigma (1-\eta (R,\{ R_{i} \} )S, \quad \eta =\frac{S_{m} (R,\{ R_{i} \} )}{S} ,}\\[0.3cm]
{\delta \varepsilon _{int} (R,\{ R_{i} \} ,\theta _{1} )=\sigma \delta S_{m} (R,\{ R_{i} \} ,\theta _{1} )}
\end{gathered}
\end{equation}

Here, $V$ is the volume of the pore, $\sigma $ is the surface energy of the liquid, $\delta \sigma =(\sigma _{ls} -\sigma _{sg} )$ is the difference between the surface energies of the solid--liquid and solid--gas interfaces, $\delta S_{m} $ is a change in the surface of menisci at the intrusion of the liquid into a pore with radius $R$, and $\eta $ is the connectivity factor. Below, we will use the mean field approximation; in this case,

\begin{equation}
\label{eq5}
\begin{gathered}
{\delta \varepsilon _{1} (R)=pV-\delta \sigma (1-\eta (R,\theta _{1} ))S, \quad \eta (R,\theta _{1} )=\frac{\langle S_{m} (R,\theta _{1} ,\{ R_{i} \} )\rangle}{S} ,}  \\[0.3cm]
{\delta \varepsilon _{int} (R,\theta _{1} )=\sigma \langle \delta S_{m} (R,\{ R_{i} \} ,\theta _{1} )\rangle}
\end{gathered}
\end{equation}

Here, $\langle \quad \rangle$ means averaging over all pores of the medium except for the given pore with radius $R$. In this approximation, the area of the menisci $S_{m} $ and its change $\delta S_{m} $ are determined by the nearest neighbors of the given pore. Therefore,

\begin{equation}
\label{eq6}
\begin{gathered}
\eta (R,\theta _{1} )=\frac{1}{4\pi R^{2} } \int _{0}^{\infty }z(R,R_{1} ,\theta _{1} )s_{m} (R,R_{1} )f(R_{1} )dR_{1}  \\[0.3cm]
\langle\delta S_{m} (R,\{ R_{i} \} ,\theta _{1} )\rangle=\int dR_{1}  f(R_{1} )z(R,R_{1} )s_{m} (R,R_{1} )W_{1} (z(R,R_{1} ),\theta _{} ).  
\end{gathered}
\end{equation}

Here, $z(R,R_{1} )$, $s_{m} (R,R_{1} )$ are the number of the nearest neighbors and the area of the meniscus of the given pore with radius $R$, respectively, and $W_{1} (z(R,R_{1} ),\theta _{1} )$ takes into account the interaction between liquid nanoclusters in neighboring pores and is the difference between the average numbers of menisci before and after the extrusion of the liquid from the pore per nearest neighbor. The product of $W_{1} (z(R,R_{1} ),\theta _{1} )$ and the surface energy of the liquid in menisci determines a possible change in the energy of the interaction of liquid nanoclusters with the environment at the transition to the unstable state. 

The extrusion of the liquid from the filled medium first occurs at a decrease in the pressure through the formation of individual empty pores and clusters of empty pores near the percolation threshold on the shell of the infinite cluster of filled pores. At $\theta _{1} >\theta _{c} $, $W_{1} (\theta _{1} >\theta _{c} )$ should be determined as the difference between the numbers of menisci after and before the extrusion of the liquid from a pore on the shell of the system of empty pores:

\begin{equation}
\label{eq7}
\begin{gathered}
W_{1} (R,R_{1} ,\theta _{1} >\theta _{c})=\sum _{n=0}^{z-1}(1-\theta _{1})^{n} (P(\theta _{1} ))^{z-n} \frac{z-2n}{z} \frac{z!}{n!(z-n)!} ,  \\[0.3cm]
z=z(R,R_{1} ,\theta _{1}).  
\end{gathered}
\end{equation}

Here,$P(\theta _{1} )$ is the probability that the filled pore belongs to the infinite cluster of filled pores. The first factor is the probability that an empty pore is located near the infinite cluster of filled pores. The second factor is the probability that this pore is surrounded by $z-n$ empty pores and, hence, has $z-n$ throats. The third factor is the difference between the relative numbers of menisci after ($z-n$) and before ($n$) the intrusion of the liquid into the pore. The fourth, combinatory, factor takes into account the variants of the distribution of $n$ menisci in the nearest neighbors of this pore. Summation takes into account all possible variants of the mutual arrangement of empty and filled pores. 

It follows from Eq.~(\ref{eq7}) that $W_{1} (\theta _{1} )$ at $\theta _{1} \to 1$ tends to 1, which corresponds to a change in the number of menisci at the extrusion of the liquid from one pore in the completely filled porous medium. A decrease in the filling factor is accompanied by a decrease in $W_{1} (\theta _{1} )$ and a change in sign at $\theta _{1} =\theta _{0} (z)$, which corresponds, according to Eqs.~(\ref{eq5}) and (\ref{eq6}), to a change in the sign of the energy of the interaction of liquid nanoclusters with the environment at the transition to an unstable state. In this case, liquid nanoclusters in pores can become unstable and be extruded. At $\theta _{1} <\theta _{c} $, $W_{1} (\theta _{1} )$ is constant, which corresponds to the destruction of individual liquid nanoclusters.

The connectivity factor $\eta (R,\theta _{1} )$ and quantity $\langle\delta S_{m} \rangle$ can be calculated using a particular model of the porous medium. In this work, they are calculated using the model of randomly located spheres \cite{19} with pores of various radii taking into account correlations in the spatial arrangement of pores in the medium. In this model, 

\begin{equation}
\label{eq8} 
z(R,R_{1} )=\frac{1}{\varphi V_{pore} } \int _{\left|R-R_{1} \right|}^{\left|R+R_{1} \right|}g_{1}  (R,R_{1} ,r)\cdot d^{3} r,  
\end{equation}

where $V_{pore}$ is the volume of one pore and

\begin{equation}
\label{eq9}
\begin{gathered}
g_{1} (R,R_{1} ,r)=\varphi ^{\frac{1}{R_{1}^{3} } \left(R^{3} +R_{1}^{3} -\frac{3}{4} \cdot x^{2} (R_{1} -\frac{x}{3})-\frac{3}{4}\cdot y^{2} (R-\frac{y}{3})\right)} ,\\[0.3cm]
x=\frac{R^{2} -(r-R_{1} )^{2} }{2r}, \quad y=R+R_{1} -x-r 
\end{gathered}
\end{equation}
 
is the pair correlation function of the system of randomly located spherical pores with radii $R$ and $R_{1} $, and

\begin{equation}
\label{eq10}
\begin{gathered}
s_{m} (R,R_{1} )=\frac{1}{V} \int _{\left|R-R_{1} \right|}^{R+R_{1} }\frac{\pi R\left(R_{1}^{2} -\left(r-R\right)^{2} \right)}{r}  \cdot g_{2} (R,R_{1} ,r)4\pi r^{2} \cdot dr,\\[0.3cm]
V=\frac{4\pi}{3} \left(\left(R+R_{1} \right)^{3} -\left|R-R_{1} \right|^{3} \right). 
\end{gathered}
\end{equation}

According to calculations by Eqs.~(\ref{eq6}) and (\ref{eq8})-(\ref{eq10}), $\eta \sim (R_{0} /R)^{-\alpha } ,\alpha \approx 0.3$ in this model, where $R_{0} $ is determined by the pore size distribution function and is about the minimum radius of pores. In this case, since the pore size distribution is narrow,

\begin{equation*}
\begin{gathered}
\langle\delta S_{m} (R,\{ R_{i} \} ,\theta )\rangle=\\[0.3cm]
=\int dR_{1}  f(R_{1} )z(R,R_{1} )s_{m} (R,R_{1} )W_{1} (z(R,R_{1} ),\theta )\approx \eta (R)W_{1} (\bar{z},\theta ),
\end{gathered}
\end{equation*}
where $\bar{z}$ is the average number of the nearest neighbors in the model of randomly located spheres, which depends only on porosity: $\bar{z}=-8\ln (1-\varphi )$ \cite{19}. 

It follows from Eqs.~(\ref{eq5}) and (\ref{eq6}) that a change in the energy of the pore $\Delta E(\theta ,R)=\delta \varepsilon _{1} (R,\theta _{1} )+\delta \varepsilon _{int} (R,\theta _{1} )$ after the extrusion of the liquid from it in the partially filled porous body depends on the radius of the pore $R$ and the filling factor $\theta _{1} $. 

The possibility of the existence of liquid nanoclusters in pores in the stable state after removal of the overpressure is determined by the fluctuation probability $w$ of entrapment of the liquid in the pore \cite{15}. Since a ``quantum'' of quasistatic filling is filling of one pore and, therefore, a pore can be in one of two possible states, filled or empty, we have

\begin{equation}
\label{eq11} 
w=(1+\exp (-\frac{\Delta E(\theta ,\eta (R),R)}{T} ))^{-1} .    
\end{equation} 

If $\Delta E\left(\theta ,\eta (R),R\right)>0$, the probability is $w \sim 1$ and the liquid nanocluster in the pore is stable, whereas if $\Delta E\left(\theta ,\eta (R),R\right)<0$, the probability is $w=0$ and the liquid is extruded from the pore. At the filling factor $\theta >\theta _{c0} $, entrapment of the liquid is possible in pores with radii smaller than the critical value $R^{*} (\theta _{1} ,T)=R_{0} (1+\frac{\sigma _{\lg } }{\delta \sigma _{} } W_{1} (\bar{z},\theta _{1} ))^{\frac{1}{\alpha } } $, which is determined from the equation $\Delta E=0$ and depends on the filling factor $\theta _{1} $ and temperature $T$. Temperature dependence appears owing to the temperature dependences of the surface tension coefficients $\sigma _{\lg } ,\sigma _{ls} ,\sigma _{sg} $. The volume $\Delta V(\theta _{1} )$ of the liquid confined in the porous medium filled to the filling factor $\theta _{1} $ is determined by the integral of the distribution function $f(R)$:

\begin{equation}
\label{eq12} 
\Delta V(\theta _{1} )=\int _{0}^{R^{*} (\theta _{1} ,T)}R^{3} f(R)dR .  
\end{equation} 

Further, we calculate the volume of the remaining liquid determined by the pore space geometry. At the filling factor $\theta <1$, the probability $W_{v} (\theta _{1} )$ that the pore belongs to a geometrical cluster of filled pores that is not connected with the infinite cluster of such pores is determined by the product of the probability $\theta _{1} -P(\theta _{1} )$ that the pore does not belong to the infinite cluster of filled pores and the probability $1-\theta _{1} $ that the surrounding pores are empty. This condition guarantees the absence of paths for the extrusion of the liquid. If the number of the nearest neighbors in the porous medium is $z$, this probability is given by the following expression, where all possible geometrical configurations of clusters of $n$ filled pores are taken into account:

\begin{equation}
\label{eq13} 
W_{v} (\theta _{1} )=\sum _{n=1}^{z}(\theta _{1} -P(\theta _{1} ) )^{n} (1-\theta _{1} )^{z-n} \frac{z!}{zn!(z-n)!} =\frac{(1-\theta _{1} )^{z} -(1-P(\theta _{1} )^{z} }{z} .  
\end{equation} 

Here, all possible configurations of the arrangement of filled and empty neighboring pores are taken into account. The total volume of the liquid remaining in these clusters is determined by the fraction of the pores $\Delta N_{1} (\theta _{1} )$ located in the clusters in which the liquid remains because possible paths for extrusion are absent: $\Delta N_{1} =\int _{0}^{\theta _{1} }d\theta W_{v} (\theta ) $. According to Eq.~(\ref{eq13}), $\Delta N_{1} (\theta _{1} )$ is independent of the temperature and, above the percolation threshold (when $\theta _{1} >\theta _{c} $ and $P(\theta _{1} )>0$), tends to the constant $\Delta N_{\infty } =\int _{0}^{1}d\theta W_{v} (\theta ) $, which is determined by the dependence $P(\theta _{1} )$ and the number of the nearest neighbors $z$.

Finally, the total volume of the liquid remaining in the porous body filled to the filling factor $\theta _{1} $ is given by the expression
 
\begin{equation}
\label{eq14} 
\Delta V(\theta _{1} ,T)=\int _{0}^{R^{*} (\theta _{1} ,T)}R^{3} f(R)dR +\int _{0}^{\infty }\Delta N_{1} (\theta _{1} )f(R)R^{3} dR .     
\end{equation}

Using Eq.~(\ref{eq14}), we performed calculations for the description of the experimental data obtained. The lines in Fig.~\ref{ris:image1} are the dependences of the fraction of the nonwetting liquid confined in the porous body on the filling factor at various temperatures as calculated by Eq.~(\ref{eq14}) and the temperature dependence of the fraction of the nonwetting liquid remaining in the porous body at the filling factor $\theta _{1} =1$. The calculations were performed with the parameters of the porous medium and liquid ($\varphi ,\bar{R},\delta R/R$) corresponding to the experimental values from \cite{19}. The surface tension coefficient of water and its temperature dependence were taken from \cite{24}. The surface tension coefficient of water at $T= 293$ K is 72 mJ/m$^{2}$ \cite{26}. The quantity $\delta \sigma $ and its temperature dependence were determined from the temperature dependence of the extrusion pressure using the method described in \cite{25}. At $T= 293$ K, $\delta \sigma = 10$ mJ/m$^{2}$. 

As can be seen in the figures, the resulting theoretical dependences well describe the entire set of the experimental data, which can indicate that the proposed model is applicable not only to the description of entrapment in the system studied in \cite{18}, but also to the description of the intrusion--extrusion process in other systems for revealing the structural features of a porous medium. The applicability of this model to systems of various natures and for various experimental conditions will soon be analyzed.

Further, we discuss the stability of the state of the nonwetting liquid in the porous body. To this end, we calculate the addition to the thermodynamic potential of the (nonwetting liquid--nanoporous medium) system owing to the dispersion of the liquid. This addition is obtained by averaging the energy $\Delta E(\theta ,R)=\delta \varepsilon _{1} (R,\theta _{1} )+\delta \varepsilon _{int} (R,\theta _{1} )$ over the sizes of pores taking into account the probability of entrapment of the liquid in a pore with radius $R$ given by Eq.~(\ref{eq8}) at the filling factor $\theta _{1} $ and by multiplying the resulting expression by the fraction of pores in the porous medium:

\begin{equation}
\label{eq15} 
\begin{gathered}
\Delta F(\theta _{1} ,T)=\theta _{1} \int _{0}^{\infty }w(p\le 0,T,R,\theta _{1} )f(R)R^{3} \cdot\\[0.3cm]
\cdot (\delta \sigma (1-\eta (R))S-\sigma \langle\delta S_{m} (R,\{ R_{i} \} ,\theta _{1} )\rangle)dR.
\end{gathered}
\end{equation}

Estimates of $\Delta F(\theta _{1} ,T)$ at temperatures $T=279$, 286, and 293 K show that the induced addition to the free surface energy of the system of liquid nanoclusters owing to the dispersion of the nonwetting liquid is small and, for the L23--water system, is $\Delta F(\theta _{1} =1,T=293)\approx 3\cdot 10^{-2}$ eV per pore at temperatures above $T_{c} \approx 287$ K, which corresponds to the total extrusion of the liquid from the porous body for the L23С8--water system under consideration. At lower temperatures, $\Delta F(\theta _{1} =1,T=286)\approx 0.3$ eV and $\Delta F(\theta _{1} =1,T=279)\approx 1.6$ eV. Thus, the nonwetting liquid in the nanoporous body at $T<T_{c}$ can pass to a stable dispersed state; i.e., it efficiently becomes ``wetting''. Extrusion is energetically unprofitable for those clusters of filled pores for which the number of menisci formed in the final state after the extrusion of the liquid is larger than the initial number of menisci. In this case, the extrusion of the liquid should be accompanied by an increase in the surface energy of the liquid. For this reason, entrapment can be treated as the attraction between liquid nanoclusters in the neighboring pores. A detailed analysis and comparison of the proposed model with all reported experimental data on entrapment of the liquid in porous media will be given elsewhere.

\section{ACKNOWLEDGMENTS}

This work was supported by the Ministry of Education and Science of the Russian Federation (``Scientific and scientific-pedagogical personnel of innovative Russia for 2009--2013'' Federal Goal-Oriented Program and the ``Development of Scientific Potential of Higher School for 2009--2012'' Analytical Departmental Goal-Oriented Program).

\end{document}